\documentclass[manuscript]{aastex62}

\usepackage{color}

\newcommand{\sunrise}{\textsc{Sunrise}}
\usepackage{textcomp}

\received{} \revised{} \accepted{}
\submitjournal{ApJ}

\shorttitle{MHS extrapolation}
\shortauthors{Zhu et al.} 

\begin{document}

\title{On the extrapolation of magneto-hydro-static equilibria on the sun}

\correspondingauthor{Xiaoshuai Zhu} \email{zhu@mps.mpg.de}

\author{Xiaoshuai Zhu}
\affil{Max-Planck-Institut f\"ur Sonnensystemforschung \\
Justus-von-Liebig-Weg 3, D-37077 G\"ottingen, Germany}

\author{Thomas Wiegelmann}
\affiliation{Max-Planck-Institut f\"ur Sonnensystemforschung \\
Justus-von-Liebig-Weg 3, D-37077 G\"ottingen, Germany}

\begin{abstract}

Modeling the interface region between solar photosphere and corona is challenging, because the relative importance of magnetic and plasma forces change by several orders of magnitude. While the solar corona can be modeled by the force-free assumption, we need to take care about plasma forces (pressure gradient and gravity) in photosphere and chromosphere, here within the magneto-hydro-static (MHS) model. We solve the MHS equations with the help of an optimization principle and use vector magnetogram as boundary condition. Positive pressure and density are ensured by replacing them with two new basic variables. The Lorentz force during optimization is used to update the plasma pressure on the bottom boundary, which makes the new extrapolation works even without pressure measurement on the photosphere. Our code is tested by using a linear MHS model as reference. From the detailed analyses, we find that the newly developed MHS extrapolation recovers the reference model at high accuracy. The MHS extrapolation is, however, numerically more expensive than the nonlinear force-free field (NLFFF) extrapolation and consequently one should limit their application to regions where plasma forces become important, e.g. in a layer of about 2 Mm above the photosphere.
\end{abstract}

\keywords{Sun: magnetic fields}


\section{Introduction} \label{sec:intro}

It is a challenging problem to reconstruct the magnetic field and plasma together in the solar atmosphere. Usually in the corona, magnetic field is expected to dominate over plasma because of the low plasma $\beta$ \citep{g01}. The magnetic field is then modeled by the so-called force-free assumption \citep{ws12}. However, in the photosphere and lower chromosphere, there always exists high $\beta$ regions where the pressure gradient and gravity are also important. Still under the assumption of stationary state, the more general extrapolation which takes into account the non-magnetic-force is called the magneto-hydro-static (MHS) extrapolation.

While sophisticated approaches of force-free extrapolation have been developped in the past few decades: \cite{s64}, \cite{s67} for potential field; \cite{ch77}, \cite{s78} for linear force-free field (LFFF); and \cite{s81}, \cite{w90}, \cite{wsr00}, \cite{ys00}, \cite{rak02}, \cite{wn03}, \cite{w04a}, \cite{w04b}, \cite{vkk05}, \cite{aba06}, \cite{wis06}, \cite{hw08}, \cite{jf12}, \cite{imp14}, \cite{gxk16} for nonlinear force-free field (NLFFF), much less papers addressed the MHS extrapolation.

In the generic case, the MHS equations are not soluble analytically. However, a special class of MHS equilibria can be obtained by the following ansatz:
\begin{equation}
\nabla\times \mathbf{B}=\alpha_{0}\mathbf{B}+f(z)\nabla B_{z}\times \mathbf{e_{z}}\label{eq:linmhs_current},
\end{equation}
where the first term is a field line parallel current and the second term defines the current perpendicular to the gravity \citep{l85}. For this special form of the current, the MHS equations can be solved by the separation of variables \citep{l85,l91,l92,nr99} or a Fast-Fourier Transform \citep{a81}. This is the so-called linear MHS model, which reduces to a LFFF for $f(z)=0$. \cite{ads98} modeled the magnetic field using MHS equations derived by \cite{l92}, taking into account the pressure and gravity. The parameters in the linear MHS model, $\alpha$ and $a$, are constant in the entire computational region and a scale-height of 2 Mm was used. The authors pointed out main properties of magneto-static configurations computed with this model, namely that the field aligned part of the current density contains two parts, the $\alpha {\bf B}$ term and the horizontal currents. Different from linear force-free fields, where the current density is strictly parallel to the magnetic field, this property adds some nonlinearity regarding the field aligned currents. Another interesting property pointed out by \cite{ads98} is that (using their Eq. (4)) the changes in plasma pressure (compared to the background atmosphere model) is as stronger as more vertical the field is. This property is consistent with the observation of a reduced plasma pressure in strong field regions like sunspots. We would like to point out that the linear MHS model requires global constants $a$ and $\alpha$ and this excludes strong localized concentration of electric current and Lorentz forces. While \cite{ads98} used the linear MHS configuration to model solar structures, the main emphasis of our paper is to develop and test a nonlinear MHS code, which does not have such limitations. As we are not aware of exact nonlinear MHS solutions in 3D, we, however, test the code by comparison with a linear MHS model.

For the general MHS equations, the computationally expensive numerical codes are required. Different numerical methods have been developed for this aim, e.g., \cite{gr58} solved a system of linear equations iteratively to approach the solution of nonlinear MHS equations. An advantage of the Grad-Rubin approach is that underlying mathematical problem is well posed. A disadvantage of providing certain boundary conditions (currents or $\alpha$ in NLFFF, additional pressure in MHS) is that in reality the boundary data on both footpoints are not consistent due to measurement errors. This can lead to large differences between the solutions computed from positive and negative footpoints as shown in \cite{sdm08} for the force-free approach. The Grad-Rubin method has been extended to solve the MHS equations with gravity by \cite{gw13} and \cite{gbb16}. \cite{wn06} developed an optimization principle for computing the magnetic field and plasma pressure consistently without considering gravity. The method was tested by application to a semi-analytic MHS solution which is axisymmetric. \cite{zwd13} modeled the MHS equilibria through magnetohydrodynamic (MHD) relaxation method. The method was tested by a Sun-like numerical model and H$\alpha$ fibril observation in the chromosphere \citep{zwd16}.

On the other hand, high spatial resolution data make MHS extrapolations necessary to extrapolate photospheric vector magnetograms upwards and resolve thereby the physics of the upper photosphere and chromosphere. Compared with the height, say about 2 Mm, of the non-force-free layer, the common spatial resolution of vector magnetograms (e.g., 700km for SDO/HMI) was too low to resolve it. The modeling of this thin layer, however, becomes possible with the unprecedented small pixel size of 40 km from the \sunrise/IMaX observation. \cite{wnn15}/\cite{wnn17} applied a linear MHS model to a quiet/an active region using the line-of-sight (LOS)/vector magnetogram from IMaX observation during its first/second flight in 2009/2013.

In this paper we present a more general optimization model, where the magnetic field, plasma pressure and density are computed self-consistently. The photospheric boundary (vector magnetogram, typical pressure and density in the quiet
region) is the only input of this model, which makes it applicable to real data. The basic equations are described in Section \ref{sec:BasicEquation}, a method used to update the pressure at the bottom boundary during optimization is presented in Section \ref{sec:BotPressure}, then the algorithm is presented in Section \ref{sec:algorithm}, an analytic linear MHS solution for testing the code is described in Section \ref{sec:reference_solution}, the results are presented in Section \ref{sec:results}. In Section \ref{sec:discussion} we present our conclusions and discuss some questions in the future application.

\section{Basic Equations} \label{sec:BasicEquation}

The MHS equations are given by:
\begin{eqnarray}
\frac{1}{\mu_{0}}(\nabla \times \mathbf{B})\times \mathbf{B}-\nabla p - \rho g\mathbf{\hat{z}} & = & 0, \label{eq:force_balance}\\
\nabla \cdot \mathbf{B} & = & 0, \label{divergence_free}
\end{eqnarray}
where $\mathbf{B}$, $p$, $\rho$, $g$ and $\mu_{0}$ are magnetic field, plasma pressure, plasma density, gravitational acceleration and vacuum permeability, respectively. As the gravitational acceleration changes only $0.57\%$ (from 272.407 to 273.975 $m/s^{2}$) in the 2 Mm non-force-free layer, $g$ is treated as a constant. We define the functional
\begin{equation}
L(\mathbf{B},p,\rho)=\int_{V}\omega_{a}B^{2}\Omega_{a}^{2}+\omega_{b}B^{2}\Omega_{b}^{2}dV,\label{eq:L}
\end{equation}
with
\begin{eqnarray}
\mathbf{\Omega_{a}} &=& B^{-2}\left[\frac{1}{\mu_{0}}(\nabla \times \mathbf{B})\times \mathbf{B}-\nabla p - \rho g \mathbf{\hat{z}}\right], \\
\mathbf{\Omega_{b}} &=& B^{-2}[(\nabla \cdot \mathbf{B})\mathbf{B}],
\end{eqnarray}
where $\omega_{a}$ and $\omega_{b}$ are the weighting functions with cos-profile.

The problem of solving Eq. (\ref{eq:force_balance}-\ref{divergence_free}) is replaced with following minimization problem:
\begin{eqnarray}
minimize \quad &L&(\mathbf{B},p,\rho) \\
subject \ to: p &>& 0 \label{eq:p_constraint} \\
           \rho &>& 0 \label{eq:d_constraint}
\end{eqnarray}
The constraints can be eliminated by using the variable transformation to $p$ and $\rho$
\begin{eqnarray}
p=Q^{2}, \label{eq:Q}\\
\rho=\frac{R^{2}}{gH_{s}}, \label{eq:R}
\end{eqnarray}
where pressure scale-height $H_{s}$ is a constant. $H_{s}$ and $g$ in Eq. (\ref{eq:R}) are used to make $R$ has the dimension of $B$ and $Q$. Then the above constrained optimization problem is changed to an unconstrained one:
\begin{equation}
minimize \quad L(\mathbf{B},Q,R).
\end{equation}

According to \cite{wsr00}, \cite{w04a} and \cite{wn06}, the optimization can be simply extended to solve the MHS equations with gravity. Taking the functional derivative of the functional (\ref{eq:L}) with respect to an iteration parameter $t$ leads to:
\begin{equation}
\frac{1}{2}\frac{dL}{dt}=-\int_{V}\left(\frac{\partial \mathbf{B}}{\partial t}\cdot \mathbf{\tilde{F}}+\frac{\partial Q}{\partial t}F_{1}+\frac{\partial R}{\partial t}F_{2}\right)dV-\oint_{S}\left(\frac{\partial \mathbf{B}}{\partial t}\cdot \mathbf{\tilde{G}}+\frac{\partial Q}{\partial t}G_{1}\right)dS,\label{eq:dL}
\end{equation}
where $\mathbf{\tilde{F}},\ F_{1},\ F_{2},\ \mathbf{\tilde{G}}$ and $\mathbf{G_{1}}$ are defined in Appendix \ref{sec:variables}.

If $\mathbf{B},\ p,\ \rho$ are fixed on the boundary of the computation box, $L$ can be minimized by solving the equations
\begin{equation}
\frac{\partial \mathbf{B}}{\partial t}=\mu_{1}\mathbf{\tilde{F}},\quad \frac{\partial Q}{\partial t}=\mu_{2}F_{1},\quad \frac{\partial R}{\partial t}=\mu_{3}F_{2}\label{eq:ite_bpd}
\end{equation}
iteratively. In the paper, $\mu_{1}=\mu_{2}=\mu_{3}=1$.

\section{Consistent evolution of Pressure on the boundary} \label{sec:BotPressure}

Because of the observational limitation, only the vector magnetogram on the photosphere can be used as boundary input. The weighting functions diminish the effect of the unknown top and lateral boundaries \citep{w04a}, but different from NLFFF extrapolation we need additional information regarding the plasma pressure and density on the bottom boundary.

Because the gravitational force is only in vertical direction, we derive the following simplified MHS equations on the 2D photospheric layer:
\begin{equation}
\nabla_{ph} p=\mathbf{f}_{ph} \label{eq:FB_ph},
\end{equation}
where $\mathbf{f}_{ph}$ is the 2D Lorentz force on the photosphere and $\nabla_{ph}=\mathbf{\hat{x}}\partial_{x}+\mathbf{\hat{y}}\partial_{y}$. Taken another divergence operation on both sides of Eq. (\ref{eq:FB_ph}) results in the following Poisson's equation:
\begin{equation}
\Delta_{ph} p=\nabla\cdot\mathbf{f}_{ph}\label{eq:poisson},
\end{equation}
where $\Delta_{ph}=\partial_{x}^{2}+\partial_{y}^{2}$ is the 2D Laplacian. If we knowing the Lorentz force, the pressure is determined when the pressure on the 4 edges of the bottom plane (xy-plane) is assigned; the typical pressure of quiet region can be used as the pressure on the edges if the computation box is much larger than the active region. Although we do not know the Lorentz force of the MHS equilibria to be extrapolated, we can compute it at any step during the optimization. Then an iterative approach can be designed to update the pressure on the photosphere consistently with magnetic field (detailed description of the algorithm in Section \ref{sec:algorithm}).

From another perspective, any vector field can be decomposed into curl-free and divergence-free components (Helmholtz decomposition). For the Lorentz force on the photosphere, however, it is curl-free if the stationary state is maintained. But the Lorentz force has divergence-free component during the optimization. Taking additional divergence operation to Eq. (\ref{eq:FB_ph}) extracts the curl-free component of the Lorentz force. The curl-free component of the Lorentz force determines the pressure.


So far, we used information regarding the Lorentz force during optimization to update the bottom pressure. It looks like that the density $\rho$ can be easily computed from force balance in z-direction: $\rho=\left(\frac{1}{\mu_{0}}(\nabla\times \mathbf{B})\times \mathbf{B}-\nabla p\right)_{z}/g$. However, the test shows no improvement of the results. We will further study this issue in the future. In this paper, the bottom density is uniform and fixed during optimization.

\section{Numerical Implementation} \label{sec:algorithm}

We have developed a code to compute 3D-MHS equilibria, based on the previous optimization code \citep{w04a,wn06}.

\begin{enumerate}
\item Calculate a NLFFF by using vector magnetogram.
\item Insert an isothermal gravity stratified atmosphere. The $p$ and $\rho$ on the photosphere are uniformly distributed.
\item Iterate for $\mathbf{B},\ p$ and $\rho$ by Eq. (\ref{eq:ite_bpd}). This step repeated until L reaches its minimum\label{item:ite_bpd}.
\item Update $p$ on the photosphere by solving Poisson's Eq. (\ref{eq:poisson}) with Lorentz force computed from $\mathbf{J}\times\mathbf{B}$, and repeat from step \ref{item:ite_bpd}. If $p$ is not changed for the giving tolerance, iteration stops and output $\mathbf{B},\ p$ and $\rho$.
\end{enumerate}

\section{Reference MHS solution} \label{sec:reference_solution}

\cite{l85,l91} presented a class of analytic solutions of the 3D static, magnetized atmospheres. The solutions are characterized by two parts of electric currents as described in Eq. (\ref{eq:linmhs_current}), namely the component parallel to the magnetic field and the component perpendicular to the gravitational field. Assume that $f(z)$ has the form
\begin{equation}
f(z)=a\exp^{-\kappa z},
\end{equation}
where $a$ and $\kappa$ control the magnitude and effective height of Lorentz force. Using Fourier transforming $\mathbf{B}$ with respect to $x$ and $y$, Eq. (\ref{eq:linmhs_current}) can be solved by the separation of variables
with LOS magnetogram as bottom boundary. With this magnetic structure, the pressure and density have the following distribution:
\begin{eqnarray}
p&=&p_{0}(z)-\frac{1}{2\mu_{0}}f(z)B_{z}^{2},\\
\rho&=&-\frac{1}{g}\frac{dp_{0}}{dz}+\frac{1}{\mu_{0}g}\left[\frac{df}{dz}\frac{B_{z}^{2}}{2}+f(\mathbf{B}\cdot\nabla)B_{z}\right].
\end{eqnarray}
It is apparent from the above two equations that the plane-parallel hydro-static atmosphere ($\rho_{0}=-\frac{1}{g}\frac{dp_{0}}{dz}$, $p_{0}$) is disturbed by the magnetic field. The pressure is weak in strong $B_{z}$ region with $f>0$.

To determine all the variables in the computation box, we use the following parameter set: \cite{ll90} LOS magnetogram labeled $n=m=1$, $\Phi=\frac{\pi}{4}$ and $l=0.3$ in their notation; field line parallel linear current with $\alpha=-3.0$ and non-magnetic force with $a=0.5$; $\kappa=0.02$ means the effective height of Lorentz force is 50 grids; the background atmosphere with $\rho_{0(z=0)}=9.0\times 10^{-4} kg/m^{3}$ and temperature $T_{0(z)}=$ 6000/5500/7840K at the height 0/0.5/1.28Mm (use linear interpolation to derive inter point temperature). For more sophisticated modelling of the vertical temperature profile see \cite{val81}. Then a linear MHS solution is generated in the Cartesian box (unit: Mm)
\begin{equation}
V=\left\lbrace(x,y,z)|-1.6\le x\le 1.6,\ -1.6\le y\le 1.6,\ 0\le z\le 1.28\right\rbrace.
\end{equation} All above parameters are chosen to mimic a small magnetic pole on the sun. The grid points $80\times80\times32$ are used to resolve this reference model. The grid size is 40km which is the same with \sunrise/IMaX data.

\section{Results} \label{sec:results}

We use the figures of merit introduced by \cite{sdm06} to quantify the difference between the reconstructed magnetic field $\mathbf{B}$ and the reference one $\mathbf{b}$, and supplement these with C-value between field lines, linear Pearson correlation coefficients both for the 3D and LOS integration (along the $z$ axis) of plasma pressure ($corr3D.p,\ corr2D.p$) and density ($corr3D.\rho,\ corr2D.\rho$). They are defined as:
\begin{enumerate}
\item[$\cdot$] vector correlation
\begin{equation}
C_{vec}=\displaystyle\sum_{i}\mathbf{B}_{i}\cdot \mathbf{b}_{i}/\left(\displaystyle\sum_{i}|\mathbf{B}_{i}|^2\displaystyle\sum_{i}|\mathbf{b}_{i}|^2\right)^{\frac{1}{2}},
\end{equation}
\item[$\cdot$] Cauchy-Schwarz inequality
\begin{equation}
C_{CS}=\frac{1}{N}\displaystyle\sum_{i}\frac{\mathbf{B}_{i}\cdot \mathbf{b}_{i}}{|\mathbf{B}_{i}||\mathbf{b}_{i}|},
\end{equation}
\item[$\cdot$] normalized vector error
\begin{equation}
E_{N}=\displaystyle\sum_{i}|\mathbf{B}_{i}-\mathbf{b}_{i}|/\displaystyle\sum_{i}|\mathbf{B}_{i}|,
\end{equation}
\item[$\cdot$] mean vector error
\begin{equation}
E_{M}=\frac{1}{N}\displaystyle\sum_{i}\frac{|\mathbf{B}_{i}||\mathbf{b}_{i}|}{\mathbf{B}_{i}},
\end{equation}
where N is the number of grid points in the computation box.
\item[$\cdot$] C-value
\begin{equation}
C=\frac{1}{l^{2}}\int_{0}^{l}\sqrt{\left(\bf{r}_{ref}(\tau)-\bf{r}_{extrapol}(\tau)\right)^{2}}d\tau,
\end{equation}
where C is a measure of how well the reference and extrapolated field lines agree. It is the integration along the field line (total length $l$) from the geometrical length $\tau=0$ to $\tau=l$. The C-value has been used by \cite{wn02} and \cite{wls05} to compare the magnetic field lines with observed loops.
\item[$\cdot$] Pearson's correlation coefficient
\begin{equation}
corr.=\frac{cov(Q,q)}{\sigma_{Q}\sigma_{q}},
\end{equation}
where $cov$ is the covariance, $\sigma$ is the standard deviation, Q and q are the extrapolated solution and reference model respectively.
\end{enumerate}

\subsection{Test I: all boundary conditions provided}

\begin{figure}
\plotone{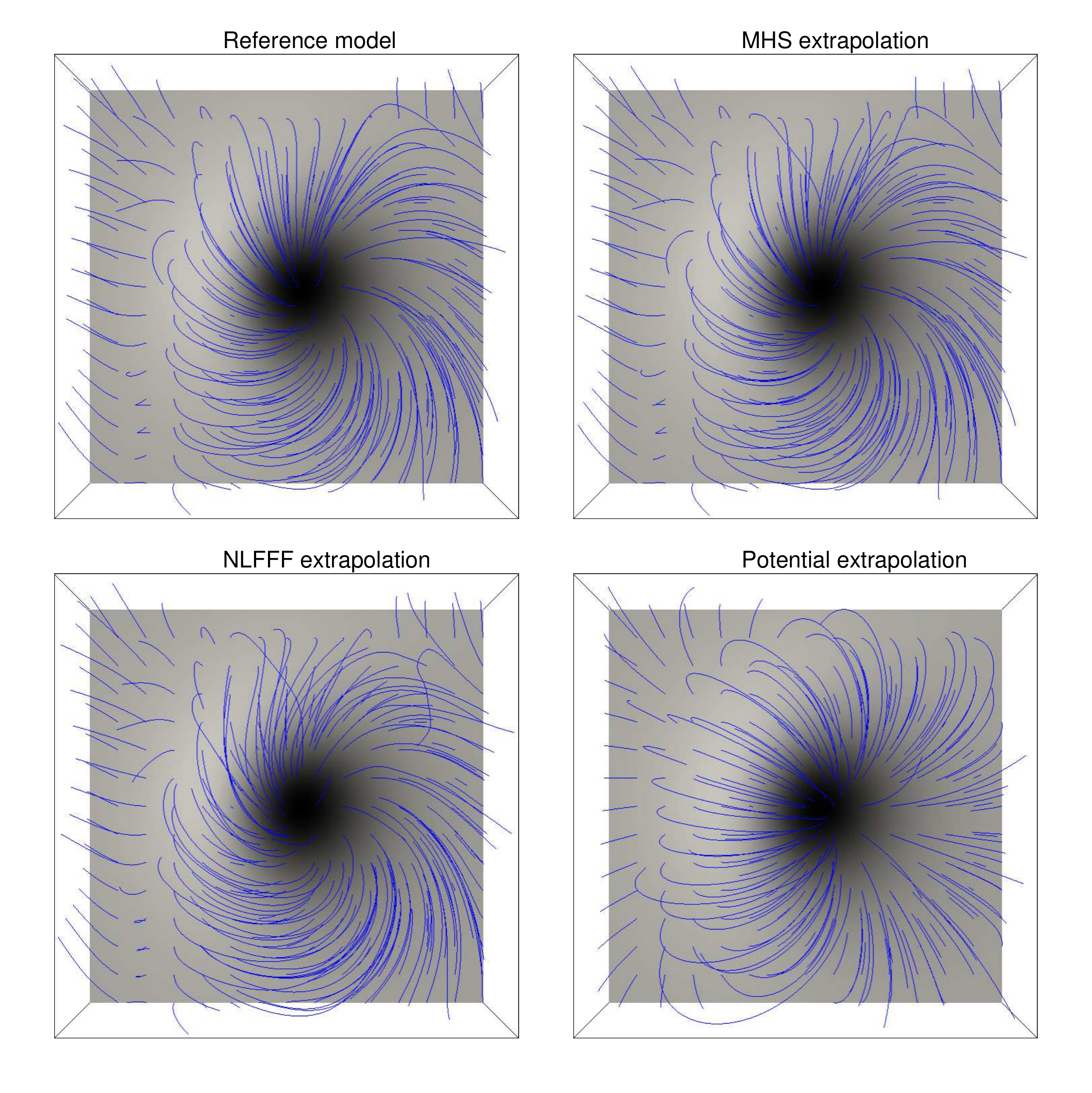}
\caption{Magnetic field for test I with different models. The field lines start from the same seeds which are uniformly distributed in the bottom plane.\label{fig:allbnd1}}
\end{figure}

\begin{figure}
\plotone{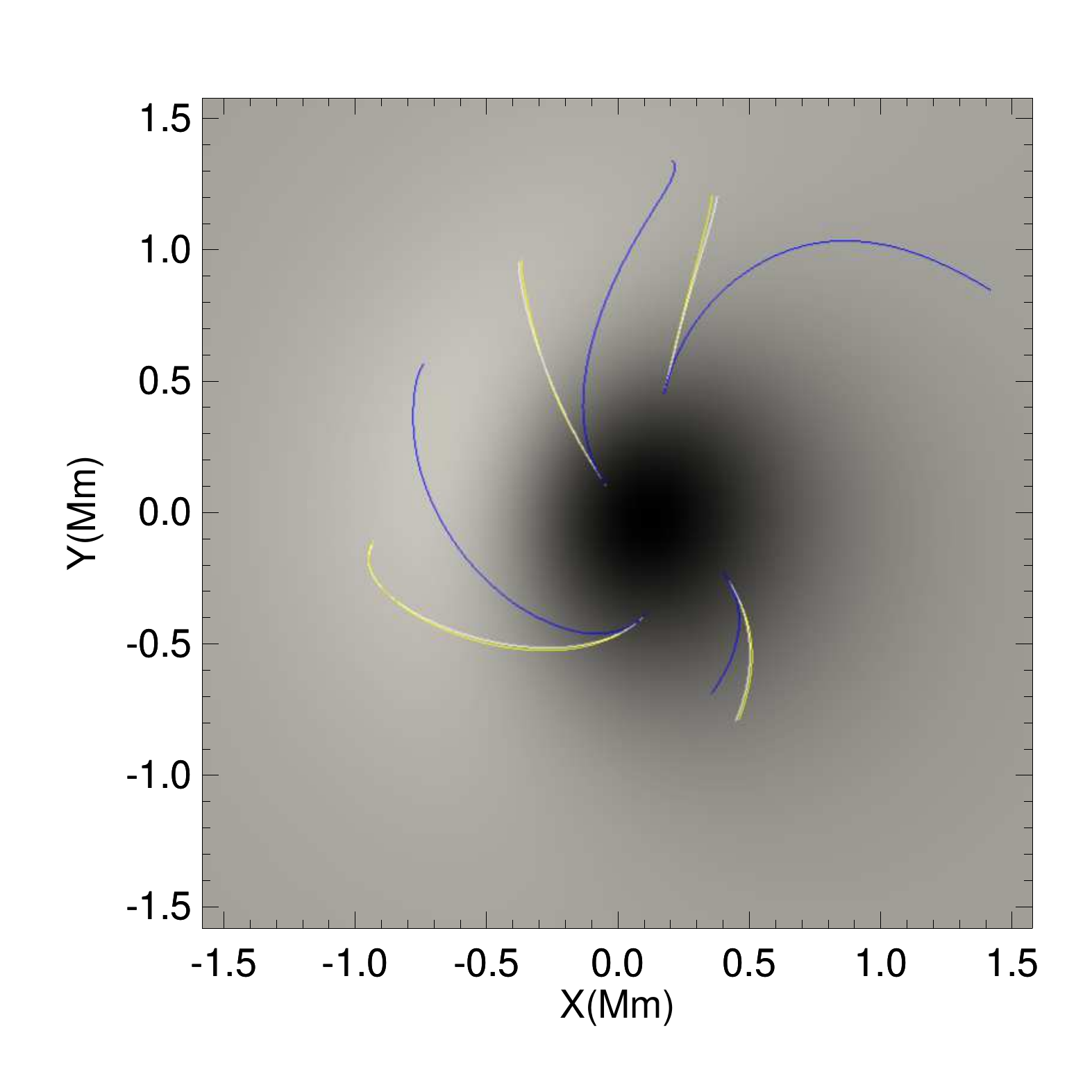}
\caption{Field lines of reference model (white), NLFFF extrapolation (blue) and MHS extrapolation (yellow) with the same start points on the bottom boundary. Notice that the white and yellow lines alsmost coincide with each other. \label{fig:allcompare}}
\end{figure}

Figure \ref{fig:allbnd1} shows the overall magnetic field line patterns from different models for test I using all boundary conditions. From Fig. \ref{fig:allcompare}, we clearly see that the MHS extrapolation produce better field lines than the NLFFF extrapolation. See also Table \ref{tab:linecompareallbnd} of the C-values of the individual field line. The mean C-values of NLFFF and MHS extrapolated lines are 0.162 and 0.016, with corresponding standard deviations of 0.128 and 0.026, respectively. The above comparisons show how the Lorentz force affects the field line patterns.

\begin{figure}
\plotone{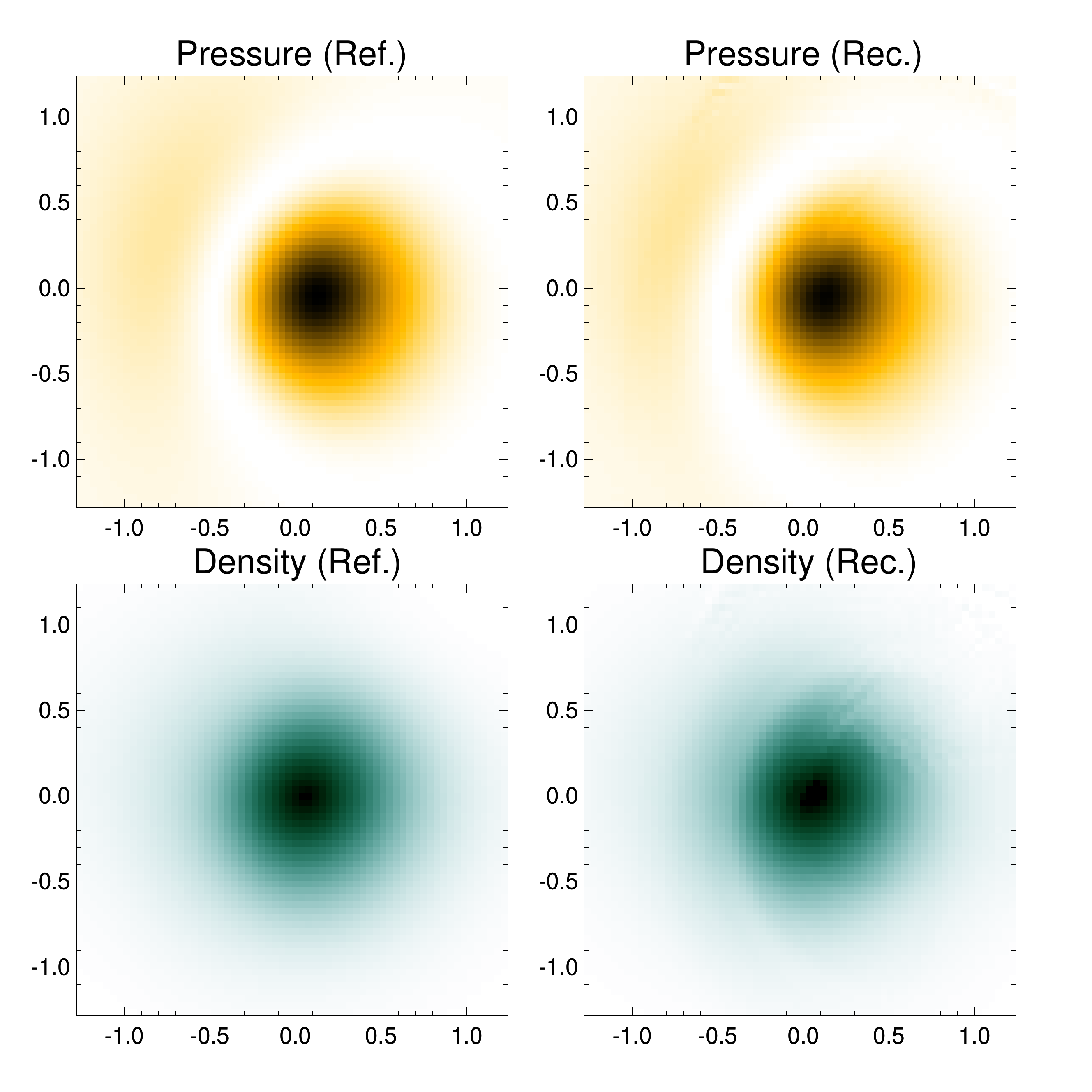}
\caption{LOS integration of the plasma pressure (top) and density (bottom) in the central field of view $x,y\in [-1.2,\ 1.2]$ (unit: Mm). Left/right panels correspond to the reference/reconstructed solution. \label{fig:allbnd2}}
\end{figure}

The ordering of the figures of merit (see Table \ref{tab:allbnd}) agrees with the conclusion from the above visual quality. Fig. \ref{fig:allbnd2} shows LOS integration of plasma pressure and density along the z-axis. We also notice that the MHS extrapolation need 5 times more steps and 6 times more CPU time than NLFFF extrapolation.

\begin{deluxetable}{cccccccccc}
\tablecaption{Model resuls for test I in which all boundary conditions are specified.}\label{tab:allbnd} \tablecolumns{10}
\tablewidth{0pt}
\tablehead{\colhead{Model}     & \colhead{$C_{vec}$} & \colhead{$C_{cs}$}   &
           \colhead{$1-E_{m}$} & \colhead{$1-E_{n}$} & \colhead{$corr2D.p$} &
           \colhead{$corr2D.\rho$} & \colhead{$corr3D.p$} & \colhead{$corr3D.\rho$} &
           \colhead{step ($\times10^{3}$)}}
\startdata
Potential & 0.8911 & 0.7841 & 0.4952 & 0.4080 & / & / & / & / & / \\
NLFFF     & 0.9875 & 0.9747 & 0.8531 & 0.8405 & / & / & / & / & 110\\
MHS       & 0.9979 & 0.9911 & 0.9492 & 0.9237 & 0.9988 & 0.9993 & 1.0000 & 0.9999 & 590\\
\enddata
\end{deluxetable}

\begin{deluxetable}{|c|cc|c|cc|c|cccc}
\tablecaption{C-values of individual field line of test I. The footpoints of 30 lines randomly distributed in the negative region ($B_{z}<0$).\label{tab:linecompareallbnd}}
\tablecolumns{9}
\tablewidth{0pt}
\tablehead{\colhead{No.} & \colhead{$C_{n\!l\!f\!\!f\!\!f}$} & \colhead{$C_{m\!h\!s}$} &
           \colhead{No.} & \colhead{$C_{n\!l\!f\!\!f\!\!f}$} & \colhead{$C_{m\!h\!s}$} &
           \colhead{No.} & \colhead{$C_{n\!l\!f\!\!f\!\!f}$} & \colhead{$C_{m\!h\!s}$} &
           \colhead{$<\!C_{n\!l\!f\!\!f\!\!f}\!>\!\pm\sigma $} & \colhead{$<\!C_{m\!h\!s}\!>\!\pm\sigma$}}
\startdata
1  & 0.059 & 0.007 & 11 & 0.277 & 0.009 & 21 & 0.003 & 0.003 & 0.162$\pm$0.128 & 0.016$\pm$0.026\\
2  & 0.105 & 0.011 & 12 & 0.103 & 0.012 & 22 & 0.025 & 0.003 \\
3  & 0.131 & 0.016 & 13 & 0.191 & 0.009 & 23 & 0.189 & 0.024 \\
4  & 0.104 & 0.017 & 14 & 0.164 & 0.006 & 24 & 0.393 & 0.076 \\
5  & 0.055 & 0.019 & 15 & 0.248 & 0.010 & 25 & 0.084 & 0.005 \\
6  & 0.013 & 0.001 & 16 & 0.234 & 0.013 & 26 & 0.035 & 0.005 \\
7  & 0.214 & 0.006 & 17 & 0.133 & 0.012 & 27 & 0.026 & 0.006 \\
8  & 0.255 & 0.012 & 18 & 0.520 & 0.014 & 28 & 0.031 & 0.007 \\
9  & 0.250 & 0.012 & 19 & 0.221 & 0.009 & 29 & 0.392 & 0.138 \\
10 & 0.262 & 0.012 & 20 & 0.025 & 0.003 & 30 & 0.118 & 0.015 \\
\enddata
\end{deluxetable}

\subsection{Test II: bottom vector magnetogram with weighted boundary layer}

In test II, we only use the bottom vector magnetogram as the boundary input, which mimics the real situation. In this test, the totally $80\times80\times32$ grids of the box consist of the inner region ($64\times64\times24$) and layer ($nd=8$ grids) at the lateral and top boundaries with cos-profile weighting functions \citep{w04a}. To see if the
pressure update on the photosphere improves the result, we perform two test runs for MHS extrapolation. The difference between them is: in one of the runs, the pressure is uniform and fixed on the photosphere during optimization; while in the other run, we update the pressure using the method mentioned in Section \ref{sec:BotPressure}.

\begin{figure}
\plotone{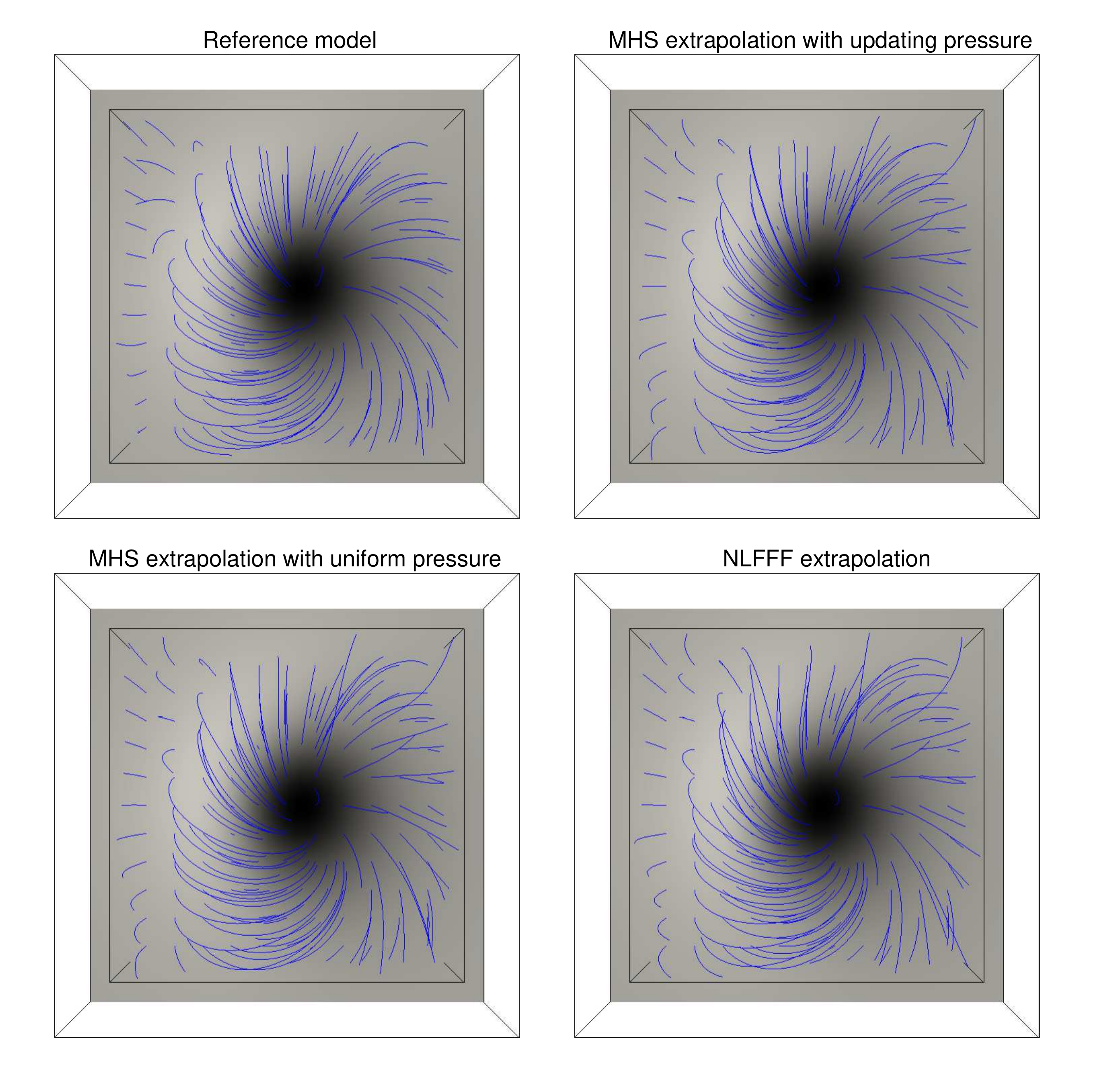}
\caption{Magnetic field in the inner region (smaller box) for test II with different models. \label{fig:botbnd1}}
\end{figure}

\begin{figure}
\plotone{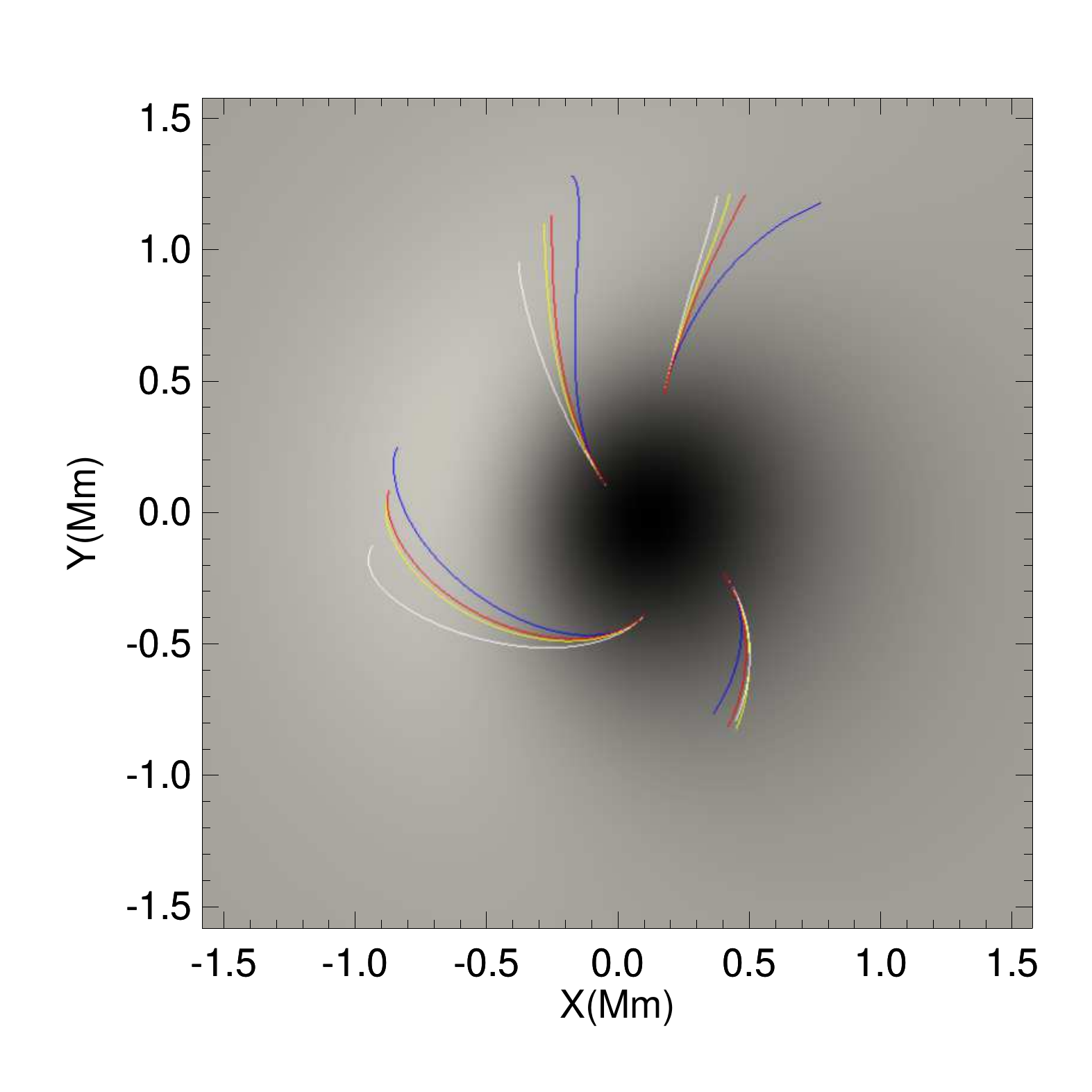}
\caption{The same with Fig. \ref{fig:allcompare} except red/yellow lines represent the magnetic field from MHS extrapolation with uniform/update bottom pressure. \label{fig:botcompare}}
\end{figure}

\begin{figure}
\plotone{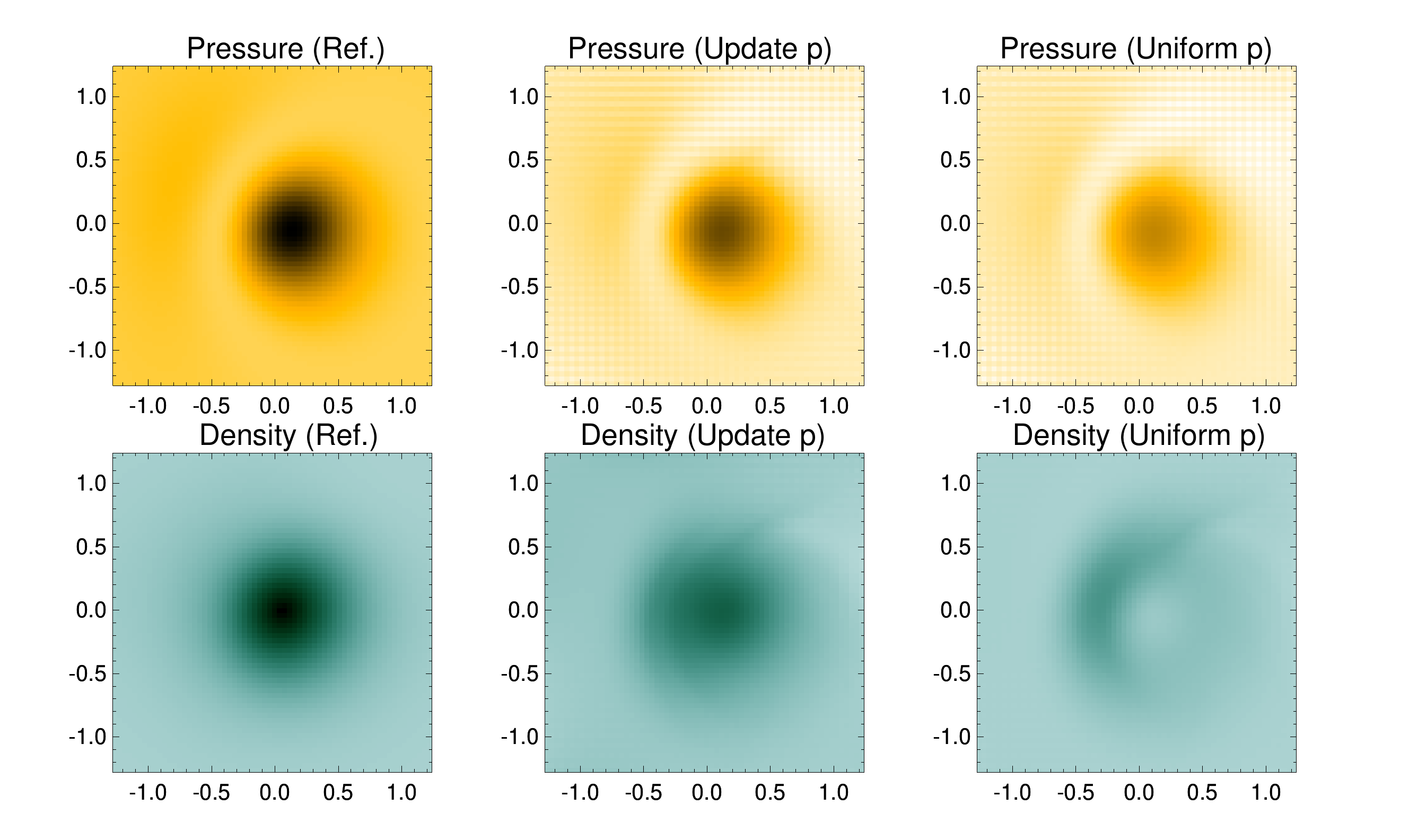}
\caption{LOS integration of the plasma pressure (top) and density (bottom) in the central field of view $x,y\in [-1.2,\ 1.2]$ (unit: Mm). Left panels are the reference results. Middle/Right panels correspond to the reconstructed solutions with updating/uniform bottom pressure.
\label{fig:botbnd2}}
\end{figure}

Figure \ref{fig:botbnd1} shows the overall magnetic field line patterns from different models for test II using bottom magnetogram. From Fig. \ref{fig:botcompare}, we can see that the MHS extrapolation produce better field lines than the NLFFF extrapolation. See also Table \ref{tab:linecomparebotbnd} of the C-values of the individual field line. The mean C-values of NLFFF and MHS extrapolated lines are 0.103 and 0.059, with corresponding standard deviations of 0.061 and 0.049, respectively. Although the field line geometry difference between the two MHS extrapolations is not large, the integration of plasma pressure and density along z-axis (see Fig. \ref{fig:botbnd2}) shows rather large differences. Updating the bottom pressure significantly improves the pressure and density results. Table \ref{tab:botbnd} shows that the ordering of the figures of merit agrees with the previous visual judgment. We notice that, for pressure and density, the correlation of 2D integration is a better index than correlation of 3D distribution. Because in 3D, the almost gravity stratified atmosphere ensures the high correlation between the reference model and reconstructed solution. It is also good to see the improvement in density result even we do not use the density information on the bottom boundary.

\begin{figure}
\plotone{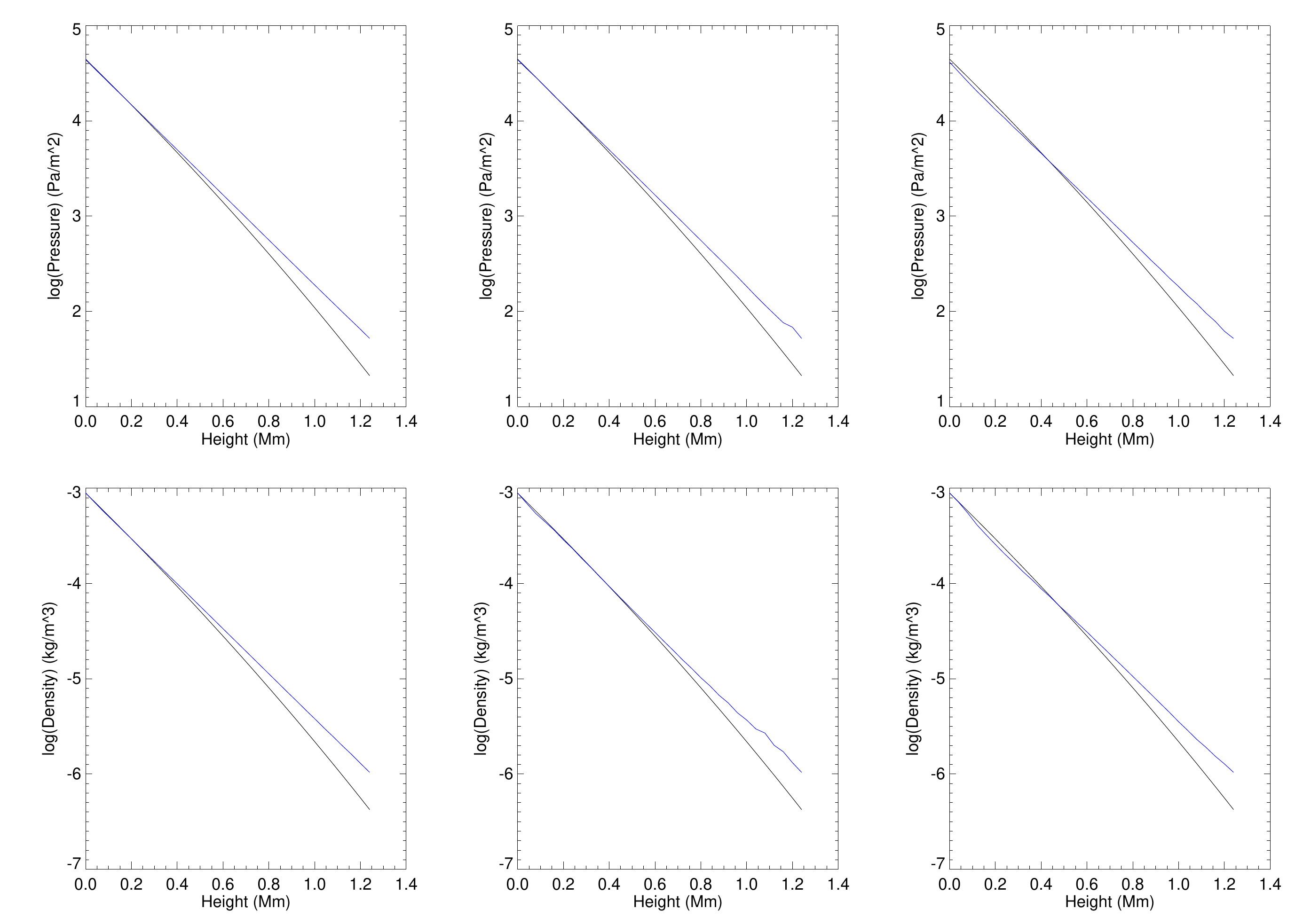}
\caption{Plasma pressure (top) and density (bottom) change along vertical axis of test II. Top/bottom panels show (from left to right): plane average pressure/density along z-axis, pressure/density along $(x,y,z)=(-0.8,\ 0,\ *)$ and $(x,y,z)=(0.8,\ 0,\ *)$. Black and blue lines correspond to the gravity stratified and final solution of the atmosphere, respectively.
\label{fig:PreDen}}
\end{figure}

\begin{figure}
\plotone{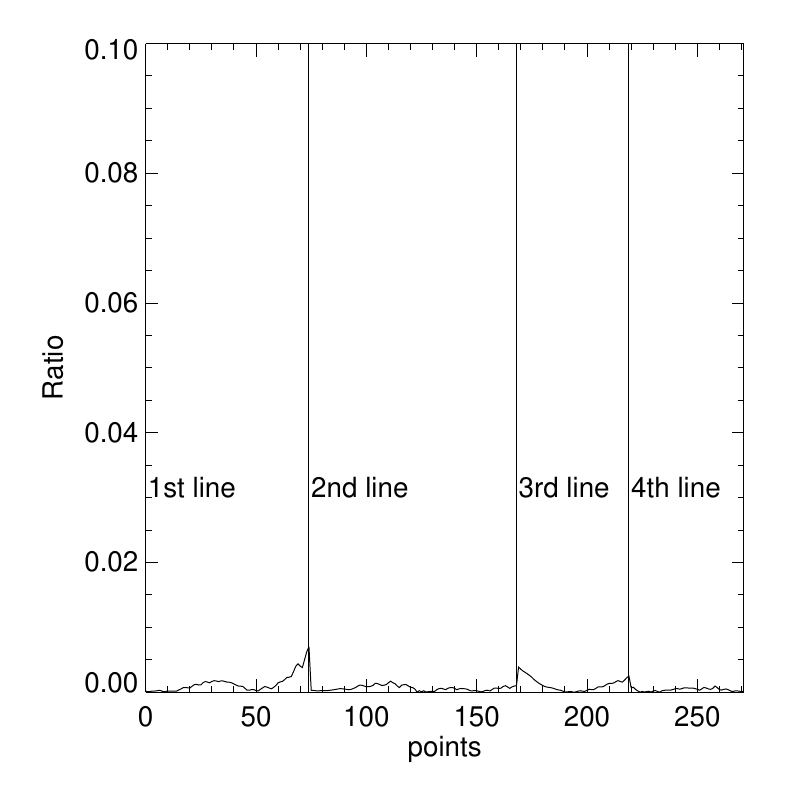}
\caption{$Ratio$ along 4 field lines (the same lines in Fig. \ref{fig:botcompare}) with totally 272 points.}
\label{fig:Ratio}
\end{figure}

The process of MHS extrapolation is optimizing the magnetic field and plasma. Fig. \ref{fig:PreDen} shows how far the final plasma deviate from the initially gravity stratified atmosphere. We can see the final solution is close to the gravity stratified atmosphere at the low height. When $z$ increases, the difference becomes larger. To check if the MHS equations are fulfilled in the extrapolated solution of test II, the field line components of $-\nabla p$ and $\rho \bf g$ are calculated. Defining
\begin{equation}
Ratio=\frac{{\bf\hat{B}}\cdot(-\nabla p+\rho\bf{g})}{|\nabla p|+|\rho {\bf g}|},
\end{equation}
where ${\bf\hat{B}}={\bf B}/B$ is the unit vector along the magnetic field line. For an MHS equilibrium, $Ratio=0$ at anywhere. Here we compute $Ratio$ along four field lines (the same lines in Fig. \ref{fig:botcompare}). For totally 272 points, the mean $Ratio$ is 0.86\textperthousand\ with standard deviation of 0.96\textperthousand\ (see Fig. \ref{fig:Ratio}). The extremely small $Ratio$ means the recovered plasma satisfy the field line component of the MHS equation at high accuracy.

\begin{deluxetable*}{cccccccccc}
\tablecaption{Results of inner region for test II with only bottom vector magnetogram specified. \label{tab:botbnd}} \tablecolumns{10}
\tablewidth{0pt}
\tablehead{\colhead{Model}     & \colhead{$C_{vec}$} & \colhead{$C_{cs}$}   &
           \colhead{$1-E_{m}$} & \colhead{$1-E_{n}$} & \colhead{$corr2D.p$} &
           \colhead{$corr2D.\rho$} & \colhead{$corr3D.p$} & \colhead{$corr3D.\rho$} &
           \colhead{step ($\times10^{3}$)}}
\startdata
Initial state\tablenotemark{a} & 0.9880 & 0.9679 & 0.8052 & 0.7538 & 0.0000 & 0.0000 & 0.9979 & 0.9978 & 6\\
Uniform-$p$\tablenotemark{b}   & 0.9917 & 0.9724 & 0.8505 & 0.7916 & 0.9694 & 0.6695 & 0.9994 & 0.9957 & 153\\
Update-$p$\tablenotemark{c}    & 0.9921 & 0.9728 & 0.8596 & 0.7977 & 0.9831 & 0.9683 & 0.9998 & 0.9982 & 191\\
\enddata
\tablenotetext{a}{Initial state consists of a NLFFF and an isothermal gravity stratified atmosphere.}
\tablenotetext{b}{Uniform bottom pressure during optimization.}
\tablenotetext{c}{Update bottom pressure during optimization.}
\end{deluxetable*}

\begin{deluxetable}{|c|cc|c|cc|c|cccc}
\tablecaption{The same with Table \ref{tab:linecompareallbnd} of test II. \label{tab:linecomparebotbnd}} \tablecolumns{9}
\tablewidth{0pt}
\tablehead{\colhead{No.} & \colhead{$C_{n\!l\!f\!\!f\!\!f}$} & \colhead{$C_{m\!h\!s}$\tablenotemark{a}} &
           \colhead{No.} & \colhead{$C_{n\!l\!f\!\!f\!\!f}$} & \colhead{$C_{m\!h\!s}$} &
           \colhead{No.} & \colhead{$C_{n\!l\!f\!\!f\!\!f}$} & \colhead{$C_{m\!h\!s}$} &
           \colhead{$<\!C_{n\!l\!f\!\!f\!\!f}\!>\!\pm\sigma $} & \colhead{$<\!C_{m\!h\!s}\!>\!\pm\sigma$}}
\startdata
1  & 0.058 & 0.011 & 11 & 0.176 & 0.054 & 21 & 0.039 & 0.036 & 0.103$\pm$0.061 & 0.059$\pm$0.049\\
2  & 0.106 & 0.025 & 12 & 0.100 & 0.018 & 22 & 0.048 & 0.047 \\
3  & 0.133 & 0.029 & 13 & 0.092 & 0.043 & 23 & 0.195 & 0.193 \\
4  & 0.116 & 0.025 & 14 & 0.039 & 0.051 & 24 & 0.092 & 0.039 \\
5  & 0.057 & 0.015 & 15 & 0.132 & 0.119 & 25 & 0.031 & 0.050 \\
6  & 0.014 & 0.010 & 16 & 0.088 & 0.065 & 26 & 0.024 & 0.017 \\
7  & 0.173 & 0.058 & 17 & 0.105 & 0.044 & 27 & 0.035 & 0.019 \\
8  & 0.181 & 0.080 & 18 & 0.193 & 0.044 & 28 & 0.058 & 0.051 \\
9  & 0.148 & 0.072 & 19 & 0.033 & 0.040 & 29 & 0.155 & 0.151 \\
10 & 0.170 & 0.076 & 20 & 0.067 & 0.073 & 30 & 0.222 & 0.207 \\
\enddata
\tablenotetext{a}{MHS extrapolation with updating bottom pressure.}
\end{deluxetable}
\subsubsection{Influence of initial conditions}

Here we investigate in the dependence of the result on the choice of initial condition. The two initial magnetic field we use are (1) potential field \citep{s78} and (2) NLFFF produced by optimization code \citep{w04a}, while the two initial atmospheres are (1) isothermal atmosphere and (2) more realistic 1D model described in section \ref{sec:reference_solution}. This results in 4 combinations.


The choice of the initial magnetic field configuration has a significant influence on the resulting  magnetic field and plasma equilibrium. Similar conclusions were found in previous studies for NLFFF \citep{w04a,sdm06}: a starting state which is near to the true solution leads to a better result. That means we better use a multigrid approach to give a
better starting state, similar as used as a standard in NLFFF extrapolation. However, notice that the initial potential field results in a somewhat more accurate density solution.

Unlike the magnetic field, the choice of the initial atmosphere has negligible influence on the results. Either isothermal atmosphere or sun-like atmosphere gives almost the same solution in this test.


\begin{deluxetable*}{cccccccc}
\tablecaption{Results of inner region with different initial conditions. \label{tab:inicondition}}
\tablecolumns{7} \tablewidth{0pt} \tablehead{
\colhead{case} & \colhead{$C_{vec}$} & \colhead{$C_{cs}$} &
\colhead{$1-E_{m}$} & \colhead{$1-E_{n}$} & \colhead{$corr2D.p$} &
\colhead{$corr2D.\rho$} & } \startdata
case I\tablenotemark{a}   & 0.9921 & 0.9728 & 0.8596 & 0.7977 & 0.9831 & 0.9683\\
case II\tablenotemark{b}  & 0.9920 & 0.9727 & 0.8594 & 0.7975 & 0.9832 & 0.9695\\
case III\tablenotemark{c} & 0.9166 & 0.8648 & 0.5648 & 0.5038 & 0.9655 & 0.9749\\
case IV\tablenotemark{d}  & 0.9165 & 0.8648 & 0.5647 & 0.5038 & 0.9653 & 0.9765\\
\enddata
\tablenotetext{a}{Initial state consists of NLFFF and isothermal atmosphere.}
\tablenotetext{b}{Initial state consists of NLFFF and sun-like atmosphere.}
\tablenotetext{c}{Initial state consists of potential field and sun-like atmosphere.}
\tablenotetext{d}{Initial state consists of potential field and isothermal atmosphere.}

\end{deluxetable*}

\subsubsection{Influence of noise}

Until now, we input the magnetic field on the bottom boundary as it is known exactly. However, this is not the case when the real vector magnetogram is used. In this subsection, we study the influence of the noise of the bottom magnetic field by adding some random noise ($2\%$ in $B_{z}$, $nl$ in $B_{x}$ and $B_{y}$) to the magnetogram. $nl=5\%, 10\%, 15\%, 20\%$ are noise levels of transverse field for different test runs. The same cos-profile weighting functions and boundary layer $nd=8$ are used in these test runs.

\begin{deluxetable*}{cccccccc}
\tablecaption{Results with different noise level $nl$. \label{tab:noise}}
\tablecolumns{7} \tablewidth{0pt} \tablehead{ \colhead{noise level} &
\colhead{$C_{vec}$} & \colhead{$C_{cs}$} & \colhead{$1-E_{m}$} &
\colhead{$1-E_{n}$} & \colhead{$corr2D.p$} & \colhead{$corr2D.\rho$} & }
\startdata
No noise   & 0.9921 & 0.9728 & 0.8596 & 0.7977 & 0.9831 & 0.9683\\
5\%        & 0.9920 & 0.9729 & 0.8576 & 0.7972 & 0.9804 & 0.9535\\
10\%       & 0.9913 & 0.9722 & 0.8514 & 0.7918 & 0.9767 & 0.9236\\
15\%       & 0.9905 & 0.9715 & 0.8453 & 0.7879 & 0.9729 & 0.8851\\
20\%       & 0.9881 & 0.9688 & 0.8316 & 0.7754 & 0.9615 & 0.8223\\
\enddata

\end{deluxetable*}

Table \ref{tab:noise} shows the results. The random noise of magnetic field is independent of neighboring grids. This leads to high frequent noise of current and Lorentz force on the photosphere, which makes the extrapolation inaccurate. As a result, all metrics are getting worse with increasing noise.

\section{Discussion and conclusions} \label{sec:discussion}

In this work, we have generalized the optimization method to apply to MHS equilibria. Compare with NLFFF approach, MHS optimization confronts two new challenges: (1) how to ensure positive pressure and density; (2) how to deal with boundary pressure and density. The first problem is actually how to deal with positivity constraint in optimization. This constraint can be eliminated by the variable transformation of Eq. (\ref{eq:Q},\ref{eq:R}). The second problem is more complex because no measurement of plasma pressure and density is available. Some information, however, are included in the data of the vector magnetogram. Based on the assumption of the force balance in the bottom plane, we obtain the Poisson's Eq. (\ref{eq:poisson}) for computing pressure on the photosphere. Then we design an algorithm to update the bottom pressure consistently within the optimization procedure. In test II, we need 18 times update of bottom pressure, and most steps (153K in totally 191K) are in the first round of $L$ minimization.

We conclude from above tests: (1) The MHS equilibria are reconstructed at relatively high accuracy by the generalized optimization principle for iterating magnetic field, plasma pressure and mass density simultaneously; (2) update the bottom pressure by using Lorentz force significantly improves the results of MHS extrapolation; (3) the initial choice of magnetic field influences the final results significantly, whereas, MHS extrapolation using a NLFFF model as the initial condition produces much better results than using a potential field.

We also test our code with vanishing $f_{z}$ by setting $a=0.0$. The model with $a=0.0$ is a LFFF. As supposed, our code can recover the LFFF at almost the same accuracy with the results obtained by NLFFF approach.

Notice that the bottom density is still uniformly distributed in the current extrapolation. We would like to address this issue in future article. In test II, the MHS extrapolation takes about 6.5 CPU hours on a 2.1GHz processor. An application to IMaX vector magnetogram embedded by HMI data (about $2000\times2000$ grids, see \cite{wnn17}) needs large amount of computational resources. For practical application, we can limit calculation to the \sunrise-FOV ($936\times 936$) to reduce the computation. A multigrid approach is likely enable faster convergence with high resolution magnetograms. Furthermore, the MHS model should be restricted to the non-force-free layer (about 2 Mm above the photosphere) to reduce the computation time. In the force-free corona above, computational less expensive NLFFF extrapolations can be used.


\acknowledgments

We appreciate the very constructive comments from the anonymous referee and the inspiring discussions with Bernd Inhester. This work was supported by DFG-grant WI 3211/4-1.

\appendix

\section{Variables definitions} \label{sec:variables}

The variables in Eq. \ref{eq:dL} are defined:
\begin{eqnarray}
F_{1}&=&-2Q\nabla \cdot (\mathbf{\omega_{a}\Omega_{a}}),\\
F_{2}&=&\frac{2\omega_{a}R}{gH_{s}}\mathbf{\Omega_{a}}\cdot \mathbf{\hat{z}},\\
G_{1}&=&2\omega_{a}Q\mathbf{\Omega_{a}}\cdot\mathbf{\hat{z}},\\
\\
\mathbf{\tilde{F}}&=&\mathbf{\tilde{F}}_{a}+\mathbf{\tilde{F}}_{b},\\
\mathbf{\tilde{G}}&=&\mathbf{\tilde{G}}_{a}+\mathbf{\tilde{G}}_{b},\\
\\
\mathbf{\tilde{F}}_{a}&=&\omega_{a}\mathbf{F}_{a}+(\mathbf{\Omega}_{a}\times \mathbf{B})\times \nabla\omega_{a},\\
\mathbf{\tilde{F}}_{b}&=&\omega_{b}\mathbf{F}_{b}+(\mathbf{\Omega}_{b}\cdot  \mathbf{B}) \nabla\omega_{b},\\
\mathbf{\tilde{G}}_{a}&=&\omega_{a}\mathbf{G}_{a},\\
\mathbf{\tilde{G}}_{b}&=&\omega_{b}\mathbf{G}_{b},\\
\mathbf{F}_{a}&=&\nabla\times(\mathbf{\Omega}_{a}\times\mathbf{B})-\mathbf{\Omega}_{a}\times(\nabla\times\mathbf{B})+\Omega_{a}^{2}\mathbf{B},\\
\mathbf{F}_{b}&=&\nabla(\mathbf{\Omega}_{b}\cdot\mathbf{B})-\mathbf{\Omega}_{b}(\nabla\cdot\mathbf{B})+\Omega_{b}^{2}\mathbf{B},\\
\mathbf{G}_{a}&=&\mathbf{\hat{n}}\times(\mathbf{\Omega}_{a}\times\mathbf{B}),\\
\mathbf{G}_{b}&=-&\mathbf{\hat{n}}(\mathbf{\Omega}_{b}\cdot\mathbf{B}),
\end{eqnarray}
where $\mathbf{\hat{n}}$ is the inward unit vector on the surface S.

\end{document}